\documentclass [12pt]{article}
\usepackage{graphicx,amssymb,amsfonts,latexsym,amsmath,amsthm,times}
\usepackage{epsfig}
\usepackage{fancyhdr}
\usepackage[usenames,dvipsnames]{color}
\usepackage[english]{babel}
\numberwithin{equation}{section}

\begin{document}


\thispagestyle{plain}

\vspace*{2cm} \normalsize \centerline{\Large \bf Modelling Pattern Formation in Dip-Coating Experiments}

\vspace*{1cm}

\centerline{\bf Markus Wilczek$^{ad}$\footnote{Corresponding
author. E-mail: markuswilczek@uni-muenster.de}, Walter B. H. Tewes$^a$, Svetlana V. Gurevich$^{ade}$, Michael H. K\"opf$^b$,} \centerline{\bf Lifeng Chi$^cd$ and Uwe Thiele$^{ade}$}

\vspace*{0.5cm}

\centerline{$^a$ Institute for Theoretical Physics, University of M\"unster, 48149 M\"unster, Germany}
\centerline{$^b$ D\'epartement de Physique, \'Ecole Normale Sup\'erieure, 75005 Paris, France}
\centerline{$^c$ Physical Institute, University of M\"unster, 48149 M\"unster, Germany}
\centerline{$^d$ Center for Nonlinear Science (CeNoS), University of M\"unster, 48149 M\"unster, Germany}
\centerline{$^e$ Center for Multiscale Theory and Computation (CMTC), University of M\"unster, 48149 M\"unster, Germany}


\vspace*{1cm}

\noindent {\bf Abstract.}
We briefly review selected mathematical models that describe the dynamics of pattern formation phenomena in dip-coating and Langmuir-Blodgett transfer experiments, where solutions or suspensions are transferred onto a substrate producing patterned deposit layers with structure length from hundreds of nanometres to tens of micrometres. The models are presented with a focus on their gradient dynamics formulations that clearly shows how the dynamics is governed by particular free energy functionals and facilitates the comparison of the models. In particular, we include a discussion of models based on long-wave hydrodynamics as well as of more phenomenological models that focus on the pattern formation processes in such systems. The models and their relations are elucidated and examples of resulting patterns are discussed before we conclude with a discussion of implications of the gradient dynamics formulation and of some related open issues.
\vspace*{0.5cm}




\vspace*{1cm}

\setcounter{equation}{0}
\section{Introduction}

The deposition of complex fluids on solid substrates is a common part of various coating techniques employed for different purposes. Examples of such techniques are spin-coating, dip-coating, slot die coating and the doctor blade technique \cite{diao2014morphology}. Although the experimental processes are often easily performed and also well controllable and reproducible, a detailed theoretical description is often still lacking, in particular, if complex fluids are deposited, where by complex we denote any fluid that has some kind of inner structure, e.g., a varying density of solute molecules. As a result pattern formation phenomena that commonly occur in such experiments 
\cite{HaLi2012acie,Lars2014aj,Thie2014acis}
are only rarely understood. This is due to the fact that among other issues a proper description of the deposition of a complex fluid on a substrate necessarily involves the modelling of the motion of a three phase contact line, i.e., of a region where substrate, complex fluid and vapour meet. For simple fluids, many different approaches and models describing such a contact line exist, however, it is still a field of active discussions and development \cite{Genn1985rmp,BEIM2009rmp,SnAn2013arfm}. Accordingly, for a complex fluid a moving contact line poses additional problems and is therefore far from well understood. \par
The importance of the contact line for pattern formation processes stems from the fact that structured depositions of a complex fluid or of one of its constituents normally occur in the proximity of a receding contact line which moves with respect to the substrate. Perhaps the simplest experimental setup to study this phenomenon is a drop of a complex fluid on a substrate, where the contact line moves due to forces resulting from capillarity and wettability and as well by evaporation/condensation \cite{Ku1914kz,HaLi2012acie,Lars2014aj,Thie2014acis}. This is, e.g., also the case for the well-known coffee stain effect \cite{DBDH1997n,Deeg2000pre,HuLa2006jpcb,MGLS2011prl,BHDK2012jcis}. Although quite regular and also complex patterns can be achieved in such a set-up \cite{KGMS1999c,THTA2012jcis,HaCC2013cgd}, such a simple approach is, however, not very suitable for quantitative experiments, as important control parameters like the velocity of the contact line cannot be controlled directly but only indirectly (e.
g., through a temperature-dependent evaporation rate).\par
A simple way to externally influence (but not control) the contact line velocity is to employ a confinement that increases the (local) vapour pressure in the gas phase. An example is the so-called sphere on flat geometry, where the liquid is confined between a flat surface and a sphere on top of it \cite{xu2007evaporation}. In this geometry the liquid forms a capillary bridge between the substrate and the sphere. While the direction of the contact line movement is controlled in this set-up, the contact line speed is still only indirectly controlled via the influence on the evaporation process.\par
A technique with immense practical impact is spin coating, where a drop of the fluid is placed at the centre of a substrate which is then rotated at high speeds. The acting centrifugal forces spread the fluid into a thin homogeneous film. The resulting film thickness can be controlled in a wide range via the rotation speed, however, the subsequent evaporation or dewetting process and therefore the contact line motion itself cannot be directly controlled in this set-up. \par
Following the classification introduced in \cite{Thie2014acis}, all the aforementioned set-ups could be called \textit{passive}, as there is no direct \textit{active} control of the contact line motion. In contrast, e.g., the doctor blade technique introduces as an additional control parameter the velocity of the blade that removes excess fluid from a completely covered substrate at an adjustable thickness. For sufficiently small thicknesses of the resulting fluid layer, the contact line directly follows the velocity of the blade. A similar technique is proposed in \cite{YaSh2005afm}, where a polymer solution is deposited at the edge of a sliding glass plate above a resting plate, where both the velocity of the top plate and the distance between the two plates can be adjusted. This is also the case for slot die coating, where the liquid is deposited onto a substrate through a slot at an adjustable height above the substrate. Here the flux \textit{and} the velocity with respect to the substrate are controlled.
 Note that most active set-ups can be used to produce either liquid layers with a small thickness that evaporate fast effectively coupling the imposed velocity and the contact line velocity \textit{or} liquid layers with a large thickness. The latter evaporate slowly only after the actual coating process, making the contact line movement independent of the previous transfer velocity, effectively making it a passive set-up.\par
In this work we focus on another active technique, the so-called dip-coating. As the name suggests, a substrate is coated by dipping it into a trough which is filled with the complex fluid that has to be applied to the substrate. The velocity at which the substrate is withdrawn from the trough has to be typically well regulated and represents an important control parameter of the experiment. The free surface of the fluid usually forms a meniscus where the substrate and the bath meet, which for small transfer velocities determines the properties of the contact line. Similarly to the doctor-blade technique or the slot die coating, dip-coating therefore provides a direct method to control the velocity of the receding contact line. Depending on the experimental conditions, various striped and branched patterns of different heights can be formed by this technique \cite{li2010controllable,li2013growth}.\par
 \begin{figure}[htbp]
\centerline{\includegraphics[width=\textwidth]{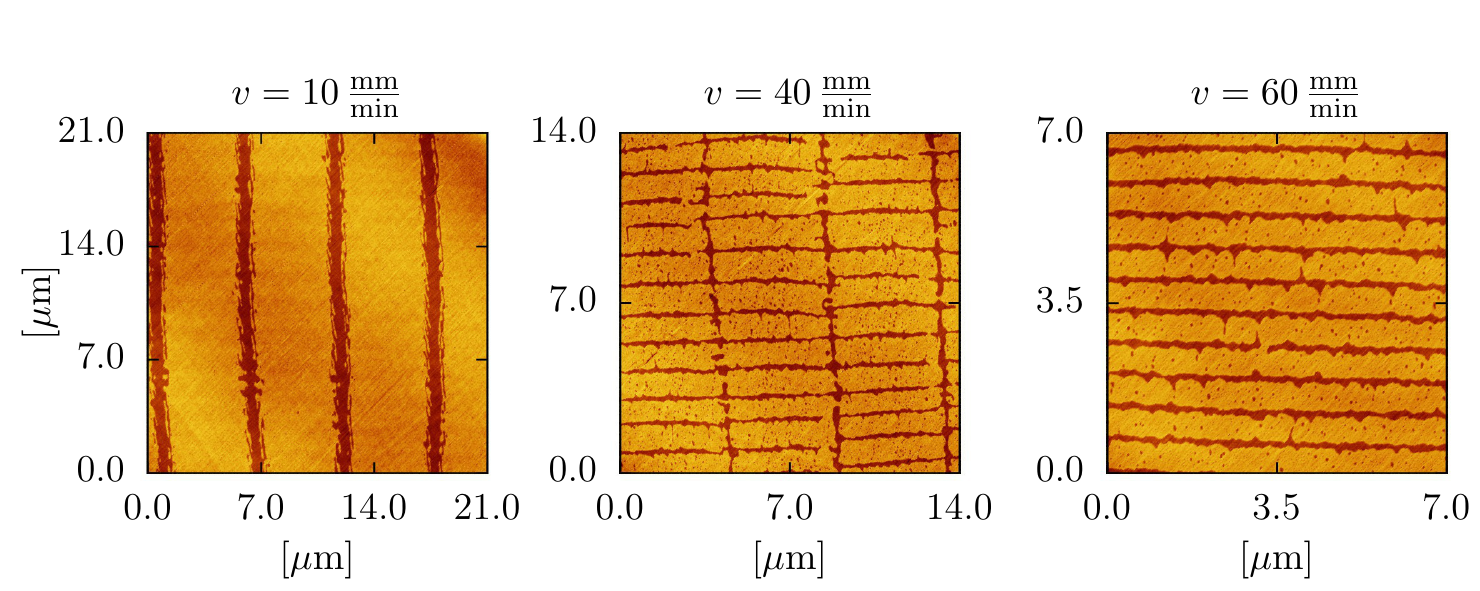}}
\caption{AFM images of a DPPC monolayer on a mica substrate. The monolayer was transferred onto the substrate by Langmuir-Blodgett transfer with different transfer velocities $v$ (direction from bottom to top) \cite{chen2007langmuir}. The different colours indicate different phases of the monolayer with different area densities and heights.}
 \label{experimentalstripes}
\end{figure}

A special case of dip-coating is the Langmuir-Blodgett (LB) transfer \cite{blodgett1935films}. For LB transfer, a trough filled with water is used, on which a floating surfactant layer is prepared. In this way a monolayer, i.e., a single layer of surfactant molecules, can be transferred onto the substrate. Due to the molecular interaction between the substrate and the surfactant film the surfactant layer can in the course of the transfer undergo a phase transition from a low density liquid expanded (LE) phase to a denser liquid condensed (LC) phase \cite{SpRi1991mcs,riegler1992structural}. 
This effect is called substrate-mediated condensation (SMC).
Detailed experimental studies show that this transition can occur in a spatially inhomogeneous way, leading to various patterns formed by domains of the two different phases \cite{SiWS1996jpc,Gleiche2000Nanoscopic,li2012structure,chen2007langmuir}. The range of resulting patterns includes stripes of different orientations and lattice-like patterns. These patterns result from self-organisation processes as their type as well as their quantitative properties (e.g., the wavelength of stripe patterns) can be controlled through experimental parameters related to out-of-equilibrium aspects as the transfer velocity of the surfactant layer or its density that is prepared at the water surface of the through. Note that, despite their striking regularity, these processes do not involve templates of any kind. Fig.~\ref{experimentalstripes} shows AFM images of a Dipalmitoylphosphatidylcholine (DPPC) monolayer on a mica substrate after LB transfer with different transfer velocities at the same surface pressure. \par
With both, dip-coating and LB transfer, regular micrometre-scale patterns can be produced across macroscopic substrate regions. As this is a desired feature in a number of production processes, a thorough theoretical understanding of the occurring processes is needed.

\vspace*{0.5cm}

The structure formation in dip-coating experiments mainly occurs at the meniscus formed at the plate when it is withdrawn from the bath. 
However, to describe the physical processes in the region of the meniscus, one needs to formulate theories for the dynamics of thin films of different complex fluids on a solid substrate. In a computational fluid dynamics approach one would describe the system 
with a full-scale macroscale deterministic continuum model, namely, employing Navier-Stokes equations 
together with transport equations for the solute(s) and accounting for solvent evaporation and processes in the gas phase. This requests a large
computational effort and suffers from uncertainties in the description of a number of processes. This approach is not reviewed here. 

However, for the considered geometries one is often able to develop asymptotic models - so-called \textit{thin film} models. Here, the attribute \textit{thin} does not necessarily refer to an absolute measure, but rather to a small ratio of typical length scales in directions perpendicular and parallel to the substrate \cite{oron1997long,CrMa2009rmp}. Often the terms \textit{long-wave approximation} and \textit{small gradient approximation} are also employed in this context. For ultrathin films one may also apply alternative approaches based on microscale considerations as, e.g., kinetic Monte Carlo models or dynamical density functional theory \cite{TVAR2009jpm,RoAT2011jpcm}. Here, however, we focus on the hydrodynamic thin film approach to the description of films of suspensions, mixtures and solutions. In particular, our brief review shall show that the various different systems of equations can be written in a unified way, namely, in a gradient dynamics form. This facilitates the comparison of the 
individual models and allows to better highlight their similarities and differences. It also allows to identify proposed models that do not comply with the gradient dynamics form and might therefore be thermodynamically inconsistent.

Thin films of simple liquids (one phase, one component) are theoretically extensively studied and various long-wave models are well established \cite{oron1997long,CrMa2009rmp,Thie2010jpcm}. Several recent studies with such models consider the dip-coating geometry and conclude that even simple fluids show a very rich behaviour of the occurring meniscus structures \cite{SADF2007jfm,SZAF2008prl,DFSA2008jfm,ZiSE2009epjt,ChSE2012pf,GTLT2014prl,TsGT2014epje}.  However, the formulation of (long-wave) models for free surface films of complex fluids on solid substrates is less advanced and a present subject of intense research and indeed debate (see, e.g., \cite{Clar2004epje,Clar2005m,NaTh2007pre,ThMF2007pf,FNOG2008pf,MaTh2009pf,TCPM2010sm,Thie11b,DoGu2013el,FAT12,BrFT2012pf,CoCl2013prl,ThTL2013prl}). For a brief review of long-wave models see \cite{Thie2014acis}.
Here we discuss the application of the underlying concepts to coating processes in active geometries \cite{koepf2010pattern,koepf2011controlled,koepf2012substrate,li2012structure,KoTh2014n,Wilczek2014Locking} and briefly set them into the context of related approaches. The particular focus is the formulation of long-wave models in the unified gradient dynamics form that allows (i) for a straightforward check of  thermodynamic consistency and (ii) for a systematic incorporation of additional physical effects.

First, we discuss in section~\ref{sec:math:general} the general form of gradient dynamics models for the evolution of a single field and the evolution of two coupled fields. Then we present two particular models for Langmuir-Blodgett transfer (section~\ref{sec:LB}) and dip-coating (section~\ref{sec:DC}) that can (i) be derived from the basic hydrodynamic transport equations and (ii) be brought into the general gradient dynamics form for two coupled fields presented before. In addition, section~\ref{sec:math:reduced} discusses a reduced model for the Langmuir-Blodgett transfer as well as possible extensions and generalisations. The work concludes in section~\ref{sec:conc} with a discussion of the relation of the presented models to related approaches.

\section{Mathematical Modelling}
\label{sec:math}

\subsection{General Form}
\label{sec:math:general}
We begin with a short discussion of the general mathematical form of the conserved parts of the evolution equations, i.e., of the dynamics of a quantity that obeys a conservation law (or continuity equation). A well-known equation for a real-valued conserved order parameter $u(\mathbf{x})$ is the so-called Cahn-Hilliard equation, that describes the dynamics of phase separation \cite{Cahn1965jcp} driven by the particular underlying free energy functional $\mathcal{F}[u]$ discussed in \cite{cahn1958free}. In its general form, it reads
\begin{align}
 \partial_t u=-\boldsymbol{\nabla}\cdot\mathbf{J}=\boldsymbol{\nabla}\cdot\left[M(u)\boldsymbol{\nabla}\frac{\delta \mathcal{F}}{\delta u}\right]
\label{eq:CH}
\end{align}
with
\begin{align}
\mathcal{F}[u]=\int\limits_{\Omega}\left[ \frac{\sigma}{2}|\boldsymbol{\nabla}u|^2+f(u)\right]~\mathrm{d}\mathbf{x}, \label{eq:CHFE}
\end{align}
where from now on we assume a two-dimensional domain $\Omega \subset \mathbb{R}^2$ and therefore $\mathbf{x} = (x, y)^\mathrm{T} \in \Omega \subset \mathbb{R}^2$ and $\boldsymbol{\nabla} = (\partial_x, \partial_y)^\mathrm{T}$, although it can also be formulated for a domain of higher dimension. 
The free energy functional consists of a local free energy $f(u)$ and an interface energy contribution $\frac{\sigma}{2}|\boldsymbol{\nabla}u|^2$. As it is suggested by our notation, the Cahn-Hilliard equation has the form of a \textit{continuity equation}, where the flux $\mathbf{J}=-M(u)\boldsymbol{\nabla}\frac{\delta \mathcal{F}}{\delta u}$ is related to a \textit{thermodynamic force}, represented by the gradient of a chemical potential $-\boldsymbol{\nabla}\frac{\delta \mathcal{F}}{\delta u}$, through the \textit{mobility function} $M(u)$. This ansatz for the flux is based on the assumption that the system is not too far from equilibrium. It can be easily seen (by calculation of $\frac{\mathrm{d}}{\mathrm{d}t} {\cal F}[u(\mathbf{x},t)]~=~\int  (\delta {\cal F}/\delta u(\mathbf{x},t))\partial_t u(\mathbf{x},t)\mathrm{d}\mathbf{x}$), that the dynamics given by equation~\eqref{eq:CH}  monotonically decreases the free energy functional $\cal F$ with time.\par
Originally, the Cahn-Hilliard equation was derived as a model of nonequilibrium thermodynamics for the evolution of binary mixtures. However, equations of such gradient dynamics form appear in many places, e.g., they turn out to be of particular interest as hydrodynamic theories of liquid films on substrates in the thin film limit \cite{Mitl1993jcis}. There they can be derived from the Navier-Stokes equations with appropriate boundary conditions by employing the \textit{long-wave approximation} \cite{oron1997long}. The time evolution of the height profile $h(\mathbf{x},t)$ of a thin liquid film on a substrate is governed by the Cahn-Hilliard-type equation
\begin{align}
 \partial_t h=\boldsymbol{\nabla}\cdot\left[\frac{h^3}{3\eta}\boldsymbol{\nabla}\frac{\delta \mathcal{F}}{\delta h}\right],\quad \mathcal{F}[h]=\int\limits_{\Omega}\left[ \frac{\sigma}{2}|\boldsymbol{\nabla}h|^2+f(h)\right]~\mathrm{d}\mathbf{x}. \label{equ2}
\end{align}
The cubic mobility function results when using no-slip boundary conditions at the substrate and stress-free conditions at the free surface and denoting the dynamic viscosity with $\eta$. In the free energy functional, the squared-gradient term represents the surface energy of the free surface of the liquid, where $\sigma$ is the liquid-gas interface tension. The local free energy $f(h)$ is identified with a potential describing the interaction energy between the free surface of the film
and the solid-liquid interface, which in the literature is referred to as disjoining potential, binding potential, adhesion potential or wetting potential, while the entire functional $F[h]$ is sometimes referred to as effective interface Hamiltonian \cite{BEIM2009rmp}. As the thin-film models in question do not contain terms due to inertia, i.e., they represent
the overdamped limit, it is rather intuitive that the dynamics minimises a free energy functional through a Cahn-Hilliard type dynamics. Nevertheless, one has to bear in mind that although the free energy functional can be derived based on thermodynamic considerations, the mobility function can only be obtained from the long-wave expansion of the basic hydrodynamic equations. \par
It is now only natural to formulate a similar model for thin films of complex fluids, which are described by $n$ conserved order parameters $\mathbf{u} = (u_1,u_2,..., u_n)^\mathrm{T}$ due to their inner structure. To describe the evolution of several order parameters governed by a single free energy functional, the Cahn-Hilliard equation~\eqref{eq:CH} can be generalised by replacing $M(u)$ with a $n \times n$ dimensional \textit{positive definite} and \textit{symmetric} mobility matrix $\mathbf{Q}(\mathbf{u})$ that encodes a thermodynamically linear coupling between all the thermodynamic forces $-\boldsymbol{\nabla}\frac{\delta \mathcal{F}}{\delta u_i}$ and the fluxes of all  order parameters. Introducing as well a symmetric $n \times n$ dimensional surface energy matrix $\mathbf{\Sigma}$ in the free energy, the equations read
\begin{align}
 \partial_t \mathbf{u}&=\boldsymbol{\nabla}\cdot\left[\mathbf{Q}(\mathbf{u})\boldsymbol{\nabla}\frac{\delta \mathcal{F}}{\delta \mathbf{u}}\right],
\label{eq:nn}\\
 \mathcal{F}[\mathbf{u}]&=\int\limits_{\Omega}\left[ \frac{1}{2} (\boldsymbol{\nabla} \mathbf{u})^\mathbf{T}\cdot \mathbf{\Sigma} (\boldsymbol{\nabla} \mathbf{u}) + f(\mathbf{u})\right] ~\mathrm{d}\mathbf{x}. \label{eq:nn2}
\end{align}
While $\mathbf{u}$ and $\frac{\delta \mathcal{F}}{\delta \mathbf{u}} = \left(\frac{\delta \mathcal{F}}{\delta u_1},\frac{\delta \mathcal{F}}{\delta u_2}, ...,\frac{\delta \mathcal{F}}{\delta u_n} \right)^\mathrm{T}$ are now $n$-dimensional vectors, one has to note that $\boldsymbol{\nabla}$ and the dot product ``$\cdot$'' are still defined on the two spatial dimensions of position space\footnote{The notation becomes clearer using an index notation with Einstein summation convention: 
\begin{align}
 \partial_t u_\alpha&=\nabla_i \left[Q_{\alpha \beta}(\mathbf{u}) \nabla_i\frac{\delta \mathcal{F}}{\delta u_\beta}\right],
\label{eq:nn3}\\
 \mathcal{F}[\mathbf{u}]&=\int\limits_{\Omega} \left[ \frac{1}{2} (\nabla_i u_\alpha)~ \Sigma_{\alpha \beta}  ~(\nabla_i u_\beta) + f(\mathbf{u})\right] ~\mathrm{d}\mathbf{x}, \quad \mathrm{with}\ i=1,2\ \mathrm{and}\ \alpha, \beta = 1,..,n. \label{eq:nn4}
\end{align}}. The constraint that the coupling of the thermodynamical forces $-\boldsymbol{\nabla}\frac{\delta \mathcal{F}}{\delta u_i}$ to all the fluxes has to occur through a positive definite and symmetric mobility matrix $\mathbf{Q}$ corresponds to the condition of positive entropy production and to the well-known \textit{Onsager reciprocity relations}, respectively \cite{Onsa1931pr,glansdorff1978thermodynamic}.\par
As the systems that we discuss next are described by two order parameters, we explicitly give the expressions~\eqref{eq:nn} and \eqref{eq:nn2} in the case $n=2$: 
\begin{align}
 \partial_t \mathbf{u}&=\boldsymbol{\nabla}\cdot\left[\mathbf{Q}(u_1,u_2)\boldsymbol{\nabla}\frac{\delta \mathcal{F}}{\delta \mathbf{u}}\right],\label{equ3}\\ \label{equ3b}
 \mathcal{F}[\mathbf{u}]&=\int\limits_{\Omega} \left[ \frac{\sigma_1}{2}|\boldsymbol{\nabla}u_1|^2+\frac{\sigma_2}{2}|\boldsymbol{\nabla}u_2|^2+ \sigma_{12} \boldsymbol{\nabla} u_1 \cdot \boldsymbol{\nabla} u_2  + f(u_1,u_2) \right] ~\mathrm{d}\mathbf{x}. 
\end{align}
An important mathematical consequence of the positive definiteness of $\mathbf{Q}$ is that the dynamics described by \eqref{equ3} still monotonically decreases the free energy $\mathcal{F}$ over time. For the two-component case one has
\begin{align}
 \frac{\mathrm{d}}{\mathrm{d}t}\mathcal{F}[u_1,u_2]&=\int\limits_{\Omega} \left(\frac{\delta \mathcal{F}}{\delta u_1}\frac{\partial u_1}{\partial t}+\frac{\delta \mathcal{F}}{\delta u_2}\frac{\partial u_2}{\partial t}\right)~\mathrm{d}\mathbf{x}
=\int\limits_{\Omega}\frac{\delta \mathcal{F}}{\delta \mathbf{u}}\partial_t\mathbf{u}~\mathrm{d}\mathbf{x}\\
&=-\int\limits_{\Omega} \left(\boldsymbol{\nabla}\frac{\delta \mathcal{F}}{\delta \mathbf{u}}\right)\cdot  \mathbf{Q}  \left(\boldsymbol{\nabla}\frac{\delta \mathcal{F}}{\delta \mathbf{u}}\right)~\mathrm{d}\mathbf{x} \leq 0.
\end{align}
In the following sections, we present two models for structure formation during dip-coating which represent models for complex liquids of the mathematical form \eqref{equ3}-\eqref{equ3b}. The presentation of the models in this unified gradient dynamics form allows for a direct comparison of the individual models and
highlights their similarities and differences. Furthermore, one may argue that for overdamped (i.e., creeping flow) interface-dominated systems it should always be possible to provide such a
gradient dynamics form that is automatically thermodynamically consistent. On this basis, on can then 
incorporate additional effects into the models in a thermodynamically consistent way
by augmenting the free energy functional with additional contributions.

\begin{figure}[htbp]
\centerline{\includegraphics[scale=0.7]{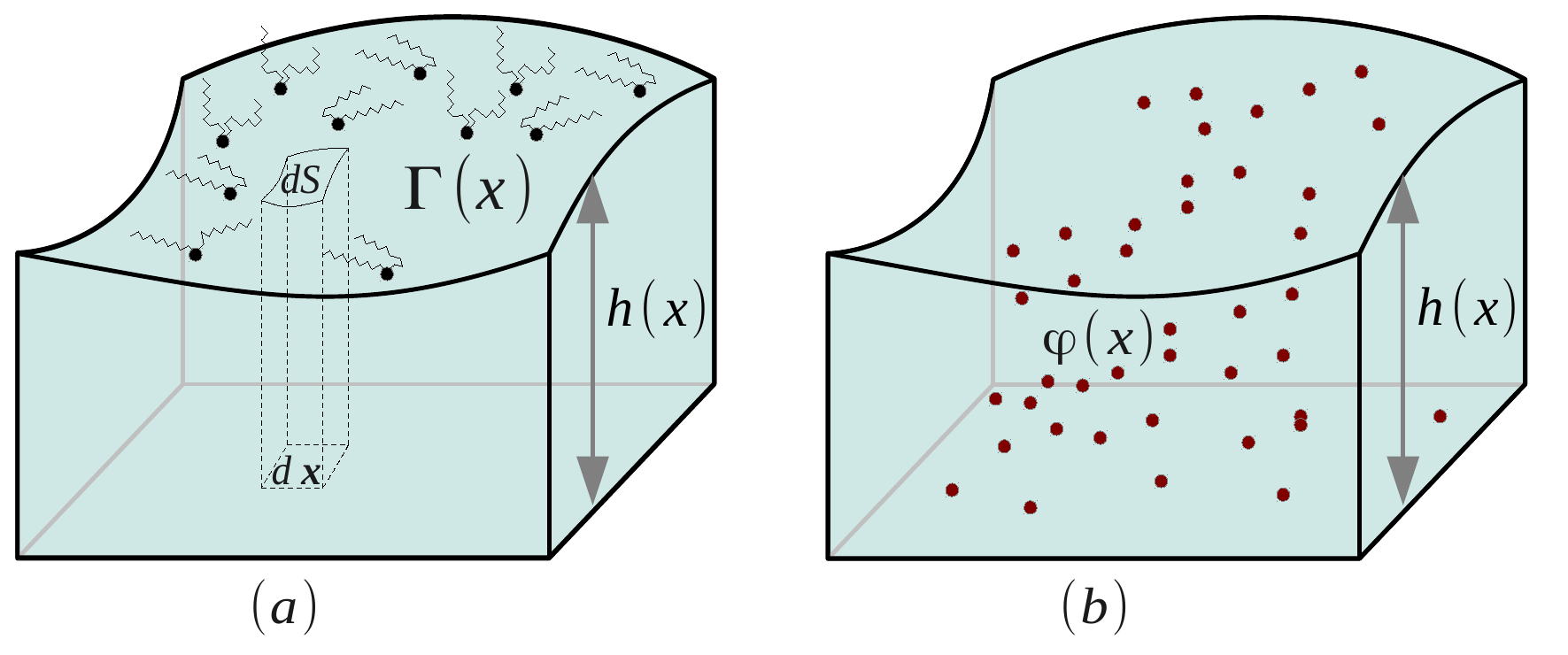}}
\caption{Sketch of the variables considered in the models for Langmuir-Blodgett transfer and dip-coating of a solution. (a): a pure liquid layer with the height $h(\mathbf{x})$ with a insoluble surfactant monolayer on top with a density $\Gamma(\mathbf{x})$, as in the Langmuir-Blodgett transfer in Section \ref{sec:LB} (b): a liquid solution layer of the height $h(\mathbf{x})$ with a solute density $\varphi(\mathbf{x})$, as in the dip-coating system in Section \ref{sec:DC}}
 \label{pic-variables}
\end{figure}

Here, we treat two applications: section~\ref{sec:LB} considers the case of a liquid layer covered by an insoluble surfactant (see Fig.~\ref{pic-variables}~(a) for a sketch and a definition of the variables), while section~\ref{sec:DC} treats films of a suspension/solution with a non-surface active solute (see Fig.~\ref{pic-variables}~(b)).

\subsection{Langmuir-Blodgett Transfer}
\label{sec:LB}
As a first example for a two order parameter gradient-dynamics theory, we discuss a model for a thin film covered by an insoluble surfactant layer \cite{ThAP2012pf}. This model can then be extended by advection and evaporation terms to model the transfer of a surfactant layer onto a substrate as it occurs in the Langmuir-Blodgett transfer described in the introduction \cite{koepf2010pattern}. \par
The natural order parameters to model a thin film covered by a surfactant are the density $\Gamma(\mathbf{x})$ of the surfactant on the surface of the liquid film and the local height $h(\mathbf{x})$ of the liquid film, see Fig.~\ref{pic-variables}~(a). However, we need to identify order parameters which are conserved on the euclidean, non-curved surface $A$ of the substrate. The physically intuitive $\Gamma$ is a density with respect to a curved film surface element $\mathrm{d}S$, i.e., $\Gamma$ and $h$ cannot be varied independently. A conserved order parameter pendant to $\Gamma$ is the density 
\begin{align}
 \tilde{\Gamma}=\xi\Gamma,\quad\mathrm{with}\ \xi=\sqrt{1+(\partial_xh)^2+(\partial_yh)^2},
\end{align}
which is the projection of the density field on the curved film surface onto the surface of the substrate, i.e., $\mathrm{d}S=\xi \mathrm{d}\mathbf{x}$. To model a system, where a surfactant phase transition may occur, we employ a local free energy that depends on $\Gamma$ and may, e.g., allow for two coexisting phases. From this local free energy follows the dependence of the surface tension of the film on surfactant density. A general free energy functional for such a system in terms of the independent conserved order parameters reads
\begin{align}
 \mathcal{F}[h,\tilde{\Gamma}]=\int\limits_{\Omega}\left[ f(h)+g\left(\frac{\tilde{\Gamma}}{\xi},h\right)\xi+\frac{\kappa}{2}\left(\boldsymbol{\nabla}\frac{\tilde{\Gamma}}{\xi}\right)^2\frac{1}{\xi}\right]\mathrm{d}\mathbf{x}.\label{equ4} 
\end{align}
In this functional, $f(h)$ is the wetting energy of the film as for the case of a simple liquid in Eq.~\eqref{equ2}. The function $g(\Gamma,h)$ is the local free energy of the surfactant layer, which allows for a phase transition which can be triggered by a change in film height. We use
\begin{equation}
 g(\Gamma,h) =\sum_{k=0}^{4} G_k \left( \Gamma - \Gamma_\mathrm{cr} \right)^k + B f(h)\left( \Gamma -\Gamma_\mathrm{cr} \right), \label{gforLB}
\end{equation}
where the $G_k$ and $B$ are constants. The first term is a double well potential, i.e., a fourth order  polynomial in $\Gamma$ centred about a critical density $\Gamma_\mathrm{cr}$ - a standard model for a first order phase transition.
Note that the zeroth order term $G_0$ corresponds to the surface energy of the liquid layer with a homogeneous surfactant coverage of density $\Gamma_\mathrm{cr}$ (denoted as $\sigma$ in Eq.~\eqref{equ2}). 
The final term in $g(\Gamma,h)$ is linear in $\left( \Gamma -\Gamma_\mathrm{cr} \right)$ and models the influence of the substrate-mediated condensation (SMC) in the form of a film-height-dependent tilt of the double well potential, that energetically favours the condensed phase when the distance between the surfactant and the substrate, i.e., the height of the liquid film, becomes small. There exist different ways to implement the tilt, in particular, as no measurements of surfactant pressure isotherms in dependence of film heights are available. It is known, however, that the SMC
lowers the free energy of the high-density LC phase for a vanishing distance between the monolayer and the substrate. Furthermore, this influence should decay fast with increasing film heights and disappear over a length scale comparable to the length scale over which the disjoining potential is acting. Therefore, here the latter's height dependence is chosen to control the tilt \cite{koepf2010pattern}, but other choices are conceivable.

The final term in Eq.~\eqref{equ4} is an energy contribution for the surfactant that penalises spatial inhomogeneities of the surfactant density, i.e., it is the energy of the one-dimensional interfaces between areas on the film surface with different surfactant phases.  The parameter $\kappa$ is a corresponding interface stiffness. Note that the various factors $\xi$ in the free energy functional $\mathcal{F}$ stem from the projection of surface densities and surface gradients onto the substrate plane.

\begin{figure}[ht]
\centerline{\includegraphics[width=0.7\textwidth]{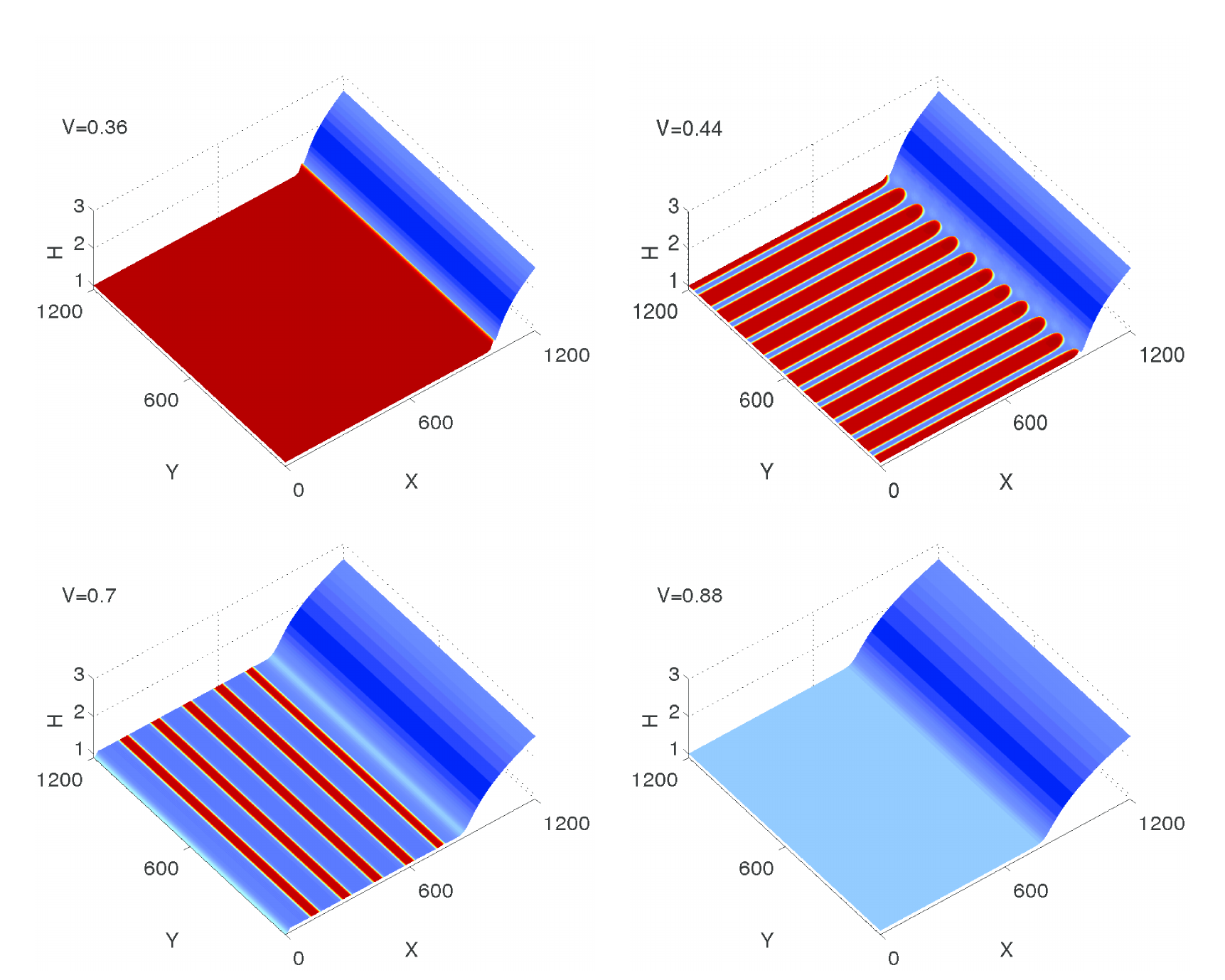}}
\caption{Snapshots from direct numerical simulations of the model for Langmuir-Blodgett transfer \eqref{equ5} \cite{koepf2010pattern,Koepf2011Phd}, where the height indicates the height of the liquid and the colour scheme encodes the density of the surfactant layer. Depending on the transfer velocity $v$, one observes the deposition of a homogeneous high-density layer, a stripe pattern orthogonal to the meniscus, a stripe pattern parallel to the meniscus, or a homogeneous low-density layer. For simulation details and parameters see \cite{Koepf2011Phd}.}
 \label{pic-LB-full}
\end{figure}

The full gradient dynamics is given by \cite{ThAP2012pf}
\begin{align}
 \partial_t h&=\boldsymbol{\nabla}\cdot\left[Q_{hh}\boldsymbol{\nabla}\frac{\delta\mathcal{F}}{\delta h}+Q_{ h \Gamma}\boldsymbol{\nabla}\frac{\delta\mathcal{F}}{\delta \tilde{\Gamma}}\right],\label{equ5}\\\nonumber
 \partial_t \tilde{\Gamma}&=\boldsymbol{\nabla}\cdot\left[Q_{\Gamma h}\boldsymbol{\nabla}\frac{\delta\mathcal{F}}{\delta h}+Q_{\Gamma \Gamma}\boldsymbol{\nabla}\frac{\delta\mathcal{F}}{\delta \tilde{\Gamma}}\right], 
 \end{align}
where the mobility matrix is
\begin{align}
\mathbf{Q}=\left(\begin{matrix} \frac{h^3}{3\eta}&\frac{h^2\Gamma}{2\eta}\\ \frac{h^2\Gamma}{2\eta}&\frac{h\Gamma^2}{\eta}+\tilde{D}\Gamma\end{matrix}\right),
\label{equ5qq}
 \end{align}
with the molecular mobility $\tilde{D}$. Note, that although the mobility matrix has always to be obtained from hydrodynamics, one may generalise the set of equations by amending or extending the free energy functional. 
The equations one obtains when performing the variations of \eqref{equ4} are (i) in the low concentration limit 
(i.e., using Eq.~(\ref{equ4}) with entropic contributions to $g(\Gamma)$ only, and with $\kappa=0$) \cite{ThAP2012pf}
identical to equations one obtains via a long-wave approximation of an advection-diffusion equation for the surfactant and the Navier-Stokes equations with appropriate boundary conditions at the substrate and at the surfactant covered surface (where a linear Marangoni effect is incorporated) \cite{JeGr1993pfa,CrMa2009rmp} and (ii) also recover the model for substrate-mediated condensation derived
in \cite{koepf2010pattern}.
There, the equations resulting from \eqref{equ5} with the free energy functional (\ref{equ4}) were directly derived based on a long-wave approximation and then supplemented with an evaporation term and an advection term to model the Langmuir-Blodgett transfer.\footnote{The resulting equations contain a term $\partial_h g(\Gamma,h)$ that was missed in the derivation presented in \cite{koepf2010pattern}, but only leads to a minor rescaling of the wetting energy. See Ref.~\cite{ThAP2012pf} for details.}
In order to incorporate the evaporation term in a fully consistent way with the free energy functional, the term
\begin{align}
 \delta_{\mathrm{ev}}=-E_v\left(\frac{\delta \mathcal{F}}{\delta h}-\mu_{\mathrm{v}}\right) \label{evap}
\end{align}
may be added to the evolution equation for $h$. This corresponds to an extension of the conserved Cahn-Hilliard type equation by a non-conserved \textit{Allen-Cahn} type contribution. The extension is based on the assumption that the evaporation is a close to equilibrium process driven by the difference of the chemical potential of the solvent layer given by $\frac{\delta \mathcal{F}}{\delta h}$ and the chemical potential $\mu_\mathrm{v}$ of the vapour phase which is assumed to be constant and homogeneous. The parameter $E_v$ is a constant that defines the time scale of the evaporation compared to the time scale of the conserved dynamics. The full resulting equations then read (cf.~\cite{koepf2010pattern,ThAP2012pf}),
  \begin{align}
    \partial_t h & = \boldsymbol{\nabla} \cdot \left[ \frac{h^3}{3\eta}  \boldsymbol{\nabla}  \left[ \partial_h f(h) - \boldsymbol{\nabla}\cdot \left( \gamma \boldsymbol{\nabla} h \right)  \right] - \frac{h^2}{2\eta}\boldsymbol{\nabla} \gamma - h  \mathbf{v}   \right] + \delta_{\mathrm{ev}},\\
    \partial_t \Gamma & = \boldsymbol{\nabla} \cdot \left[ \frac{h^2\Gamma}{2\eta}  \boldsymbol{\nabla}  \left[ \partial_h f(h) - \boldsymbol{\nabla}\cdot \left( \gamma \boldsymbol{\nabla} h \right)  \right] - \left(\frac{h\Gamma}{\eta} + \tilde{D} \right) \boldsymbol{\nabla} \gamma - \Gamma \mathbf{v}   \right],
   \end{align}
where we have introduced a generalised 'surface tension' $\gamma = g - \Gamma \partial_\Gamma g - \frac{\kappa}{2} \left( \boldsymbol{\nabla} \Gamma \right)^2 + \kappa \Gamma \boldsymbol{\nabla}^2 \Gamma$, as well as the transfer velocity $\mathbf{v}$. Note, that in \cite{koepf2010pattern} the approximation $\boldsymbol{\nabla}\cdot \left( \gamma \boldsymbol{\nabla} h \right) \approx  \gamma \boldsymbol{\nabla}^2 h$ was used. Direct numerical simulations of the model completed with suitable lateral boundary conditions (see \cite{koepf2010pattern} for details) show good qualitative agreement with the experimental results. The range of patterns that can be obtained is illustrated in Fig.~\ref{pic-LB-full}.
We emphasise that the general formulation reviewed here allows for a straightforward incorporation of additional physical effects in a thermodynamically consistent manner and to identify proposed models with ad-hoc additions, e.g., for
surfactant-concentration dependent disjoining pressures, that lack this consistency.
Several examples are discussed in \cite{ThAP2012pf}. A further example is an additional control of the pattern formation that can be achieved by the use of prestructured substrates. This can be modelled by a spatially modulated wetting energy $f(h,\mathbf{x})$ and results in the occurrence of more complex patterns including locking effects between periodic patterns and the prestructure on the substrate \cite{koepf2011controlled}.
\newpage
\subsection{Dip-coating}
\label{sec:DC}
As second example for a two order parameter gradient-dynamics model, we next discuss the case of a film of a solution of molecules or of a suspension of (nano)particles (that are assumed to be not surface active) that is transferred onto a substrate via dip-coating. Similar to the case of LB transfer described in the previous section, a thin film model for dilute solutions may be derived starting from hydrodynamic equations (Navier-Stokes, diffusion-advection equation).
At first sight, it might seem that the natural order parameters for such a model are the height $h(\mathbf{x},t)$ of the film and a height-averaged concentration of the solute,
\begin{align}
 \phi(\mathbf{x},t)=\frac{1}{h(\mathbf{x},t)}\int\limits_{0}^{h(\mathbf{x},t)}c(\mathbf{x},z,t)~\mathrm{d}z,
\end{align}
where $\mathbf{x}=(x,y)$ are again Cartesian coordinates defined on the surface of the substrate and $c(\mathbf{x},z,t)$ is the standard bulk concentration. A thin film model for the time evolution of $h$ and $\phi$ was derived in \cite{warner2003surface} and also used, e.g., in \cite{TVAR2009jpm,FAT11,FAT12}, while a gradient dynamics form in terms of variations in $h$ and $\phi$ is given in \cite{Clar2005m}. However, the fields $h$ and $\phi$ should not be varied independently as for a fixed local amount of solute a variation in height results as well in a change in $\phi$. Ref.~\cite{Clar2005m} accounts for this by employing constrained variations (for a brief discussions of problems we see with this approach see \cite{Thie11b,Thie2014acis}).\par
A simpler approach is taken in \cite{Thie11b,ThTL2013prl}, where it is argued that the natural independent order parameter fields are the film height $h(\mathbf{x},t)$ and the local amount of solute $\Psi(\mathbf{x},t)=h(\mathbf{x},t)\phi(\mathbf{x},t)$, i.e., an effective local solute 'layer height'.  Based on $h(\mathbf{x},t)$ and $\Psi(\mathbf{x},t)$ the dynamics may be formulated as a gradient dynamics based on a free energy functional that corresponds to an extended interface Hamiltonian. In terms of the conserved order parameter fields $h$ and $\Psi$, the equations for a thin film of solution/suspension read \cite{Thie11b,ThTL2013prl} \begin{align}
 \partial_t h&=\boldsymbol{\nabla}\cdot\left[Q_{hh}\boldsymbol{\nabla}\frac{\delta\mathcal{F}}{\delta h}+Q_{ h \Psi}\boldsymbol{\nabla}\frac{\delta\mathcal{F}}{\delta \Psi}\right],\label{VarForHPsi1}  \\ \nonumber
 \partial_t \Psi&=\boldsymbol{\nabla}\cdot\left[Q_{\Psi h}\boldsymbol{\nabla}\frac{\delta\mathcal{F}}{\delta h}+Q_{\Psi \Psi}\boldsymbol{\nabla}\frac{\delta\mathcal{F}}{\delta \Psi}\right], 
  \end{align}
with the mobility matrix 
\begin{align}
\mathbf{Q}=\left(\begin{matrix} \frac{h^3}{3\eta}&\frac{h^2\Psi}{3\eta}\\ \frac{h^2\Psi}{3\eta}&\frac{h\Psi^2}{3\eta}+\tilde{D}\Psi\end{matrix}\right)
\label{VarForHPsi1qq}
 \end{align}
and the free energy functional 
\begin{align}
 \mathcal{F}[h,\Psi]=\int\limits_{\Omega}\left[ f(h)+\frac{\sigma}{2}|\boldsymbol{\nabla}h|^2+hf_B\left(\frac{\Psi}{h}\right) \right]\mathrm{d}\mathbf{x}.\label{dilute}
\end{align}
Here, the first two terms are the respective wetting and surface energy as for the simple liquid film. The third term corresponds to the bulk free energy of the solute integrated over film height, i.e., it contains the concentration dependence of the free energy.
In the dilute limit (i.e., including only entropic contributions valid at low solute concentration, $f_B(\phi)\sim\phi\mathrm{ln}(\phi)$) one recovers the standard
hydrodynamic thin film equations that may then be expanded systematically by incorporating further entropic terms and interaction terms into the free energy functional, analogous to the classical approach of Cahn and Hilliard \cite{cahn1958free}, or simply using the Flory free energy for polymer solutions \cite{Flor1953}.
One may also incorporate influences of the solute on wettability \cite{Thie11b,ThTL2013prl}. A derivation of a fully solute-solvent symmetric version on a route via Onsager's variational principle \cite{Onsa1931prb} is discussed in \cite{XuTQ2015preprint}.

 \begin{figure}[htbp]
\centerline{\includegraphics[width=0.8\textwidth]{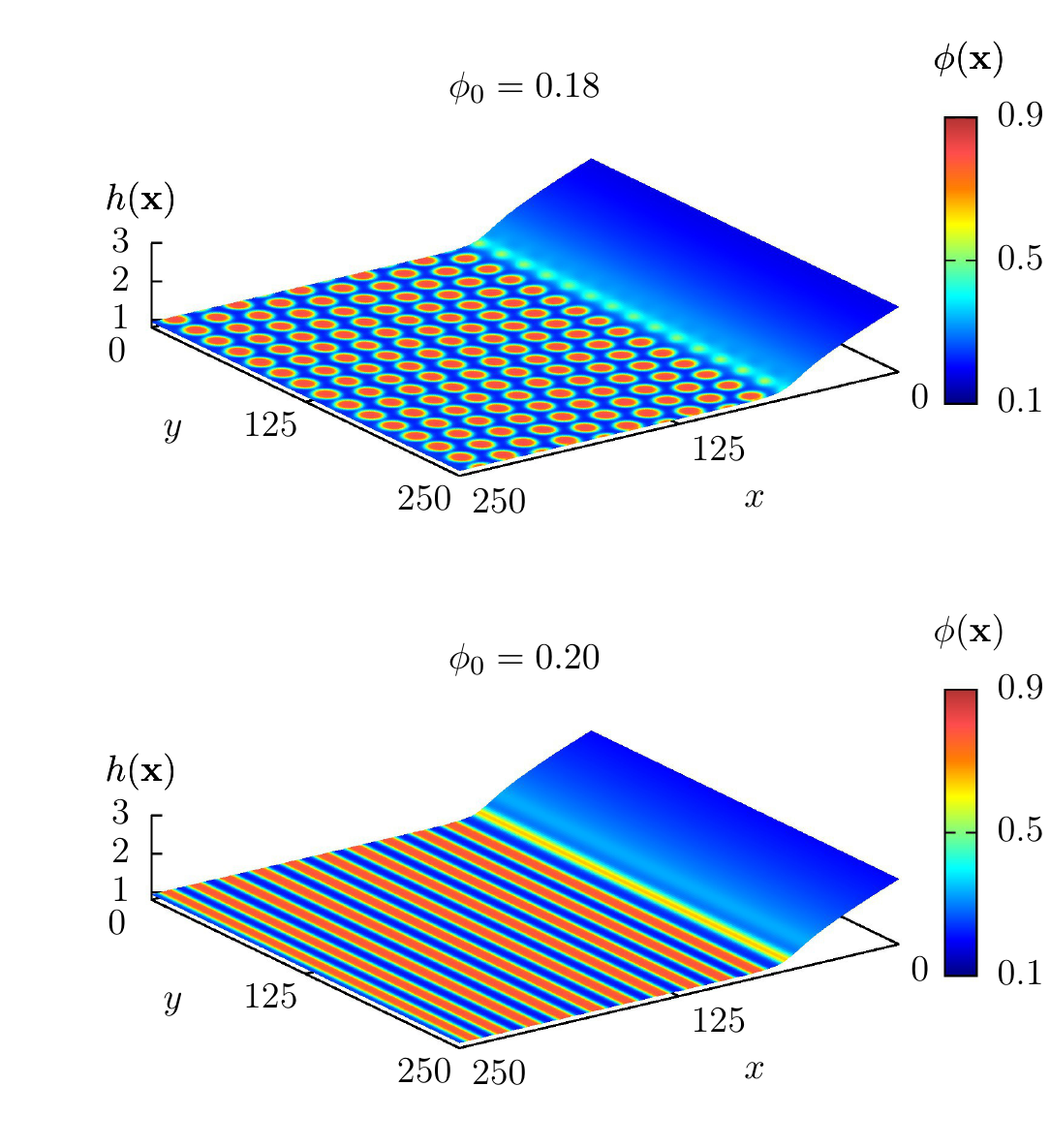}}
\caption{Snapshots from direct numerical simulations of the model for dip-coating \eqref{VarForHPsi1}, where the height indicates the height of the liquid and the colour scheme encodes the height-averaged concentration $\phi$ of the solute. Depending on the initial concentration $\phi_0$, the transfer of drops in a hexagonal pattern or stripes parallel to the meniscus is observed. For simulation details and parameters see \cite{Tewes2013msc}.}
 \label{pic-DC}
\end{figure}

As an example we consider the incorporation of entropic effects over the entire concentration range for the solute and include a Flory interactions term. Then the bulk free energy reads
\begin{align}
 f_B(\phi)=\frac{k_BT}{a^3}\left[\phi\mathrm{ln}(\phi)+(1-\phi)\mathrm{ln}(1-\phi)\right]+\alpha\phi\left(1-\phi\right)+ \frac{\kappa}{2} |\boldsymbol{\nabla}\phi|^2=\tilde{g}(\phi)+\frac{\kappa}{2} |\boldsymbol{\nabla}\phi|^2, \label{g}
\end{align}
where $a$ is a microscopic length scale related to the solute and $\alpha$ is the Flory interaction parameter.
The equations resulting from such a free energy for a general $\tilde{g}$ with the additional gradient term in $\phi(\mathbf{x},t)$ are discussed in \cite{ThTL2013prl}. The specific form in Eq.~(\ref{g}) is discussed in \cite{Tewes2013msc}.
To model the transfer process in dip-coating, we augment the gradient dynamics equations for conserved fields (\ref{VarForHPsi1}) with an appropriate evaporation term and an advection term to model the transfer due to dip-coating as done before in section~\ref{sec:LB}. In particular, evaporation results in an increased concentration within the front region of the meniscus, but should also saturate for concentrations approaching unity.
Following the same arguments as in section~\ref{sec:LB}, we choose for the evaporative contribution
\begin{align}
 \delta_{ev}=-E_v\left(\frac{\delta \mathcal{F}}{\delta h}-\mu_v\right).
 \end{align}
Evaluating the variation of $\mathcal{F}$ with respect to $h$, while keeping $\Psi$ fixed, gives $\tilde{g}-\phi\tilde{g}'$. This is a contribution of osmotic pressure type to the chemical potential of the solvent. 
Note, that without the constraint of fixed solute amount only $\tilde g'$ would result.
Osmotic pressure type contributions to evaporation are not frequent in the literature and are often introduced in an ad-hoc manner \cite{Thie2014acis}. However, in the gradient dynamics formalism it emerges naturally. For instance, when the entropic contribution $\sim(1-\phi)\mathrm{ln}(1-\phi)$ is incorporated into $\tilde{g}$, the osmotic pressure contribution
 ensures that evaporation saturates for $\phi\rightarrow 1$.
Including advection terms and appropriate lateral boundary conditions, one may study the pattern formation occurring during dip-coating experiments \cite{Tewes2013msc}. Direct numerical simulations of the presented model show various self-organised pattern types, such as, stripes parallel to the meniscus and hexagonal patterns, whose details and transitions depend on the transfer velocity and the initial concentration of the solute. Figure~\ref{pic-DC} shows two snapshots from simulations with different initial concentrations, exhibiting the two mentioned pattern types. Similar patterns were also found in \cite{YaSh2005afm}, where a related active transfer geometry is experimentally investigated.

\subsection{Reduced Models}
\label{sec:math:reduced}
\begin{figure}[htbp]
\centerline{\includegraphics[width=1\textwidth]{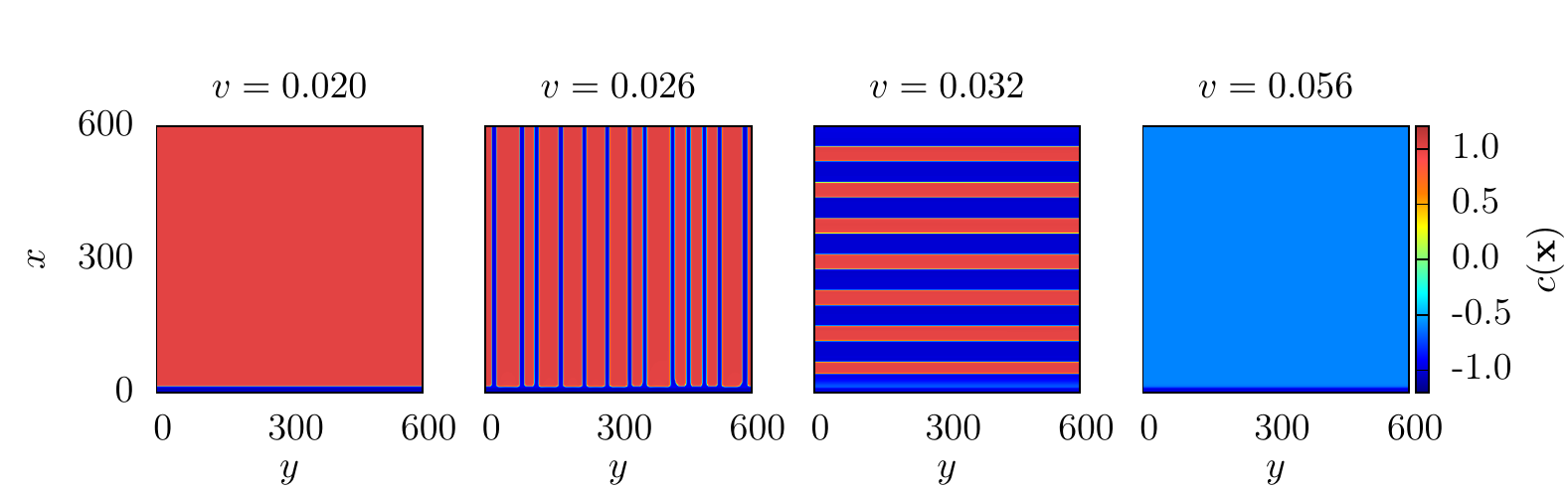}}
\caption{Different types of patterns obtained by direct numerical simulation of the reduced Cahn-Hilliard model for Langmuir-Blodgett transfer \eqref{equ6}. The range of patterns comprises stripe patterns with different orientation and adjustable wavelengths \cite{koepf2012substrate,Wilczek2014Locking}. For simulation details and parameters see \cite{Wilczek2014Locking}.}
 \label{pic-CH2D-homogeneous}
\end{figure}

Under some conditions it is justified to simplify the models presented in the previous two sections. Besides eliminating contributions to the free energy that account for a physical effect or interaction that turns out to only have a minor impact in the experimental system, one may also try to reduce the number of order parameter fields describing the system. The latter approach is feasible in the case of the model for the Langmuir-Blodgett transfer \eqref{equ5}. Time simulations of the model indicate that the film height $h(\mathbf{x},t)$ relaxes on a rather short time scale onto an almost stationary state, whose exact form has no major influence on the surfactant dynamics. Therefore it seems reasonable to replace the dynamic equation for the film height with a static approximation that enters the equation for the surfactant density $\Gamma$ only parametrically. In addition, one can show that a static height profile minimises the generalised pressure gradient, making it possible to neglect the corresponding 
term as well in the evolution equation for the surfactant density \cite{koepf2012substrate}. The resulting 
equation for the surfactant density is of simple Cahn-Hilliard type (see section~\ref{sec:math:general})
with an additional advective term. It can be written in the gradient dynamics form \eqref{eq:CH}, where we now use $c(\mathbf{x},t)$ as the order parameter for the surfactant density to avoid confusion with the full model \eqref{equ5}:
\begin{align}
 \partial_t c = \boldsymbol{\nabla} \cdot\left[M(c)\boldsymbol{\nabla}\frac{\delta \mathcal{F}}{\delta c}\right] + \mathbf{v}\cdot\boldsymbol{\nabla} c ,\quad F[c]=\int\limits_{\Omega}\left[ \frac{1}{2}|\boldsymbol{\nabla}c|^2+f(c)\right]~\mathrm{d}\mathbf{x}, \label{equ6}
\end{align}
with the mobility $M(c)$ and the local free energy $f(c)$ given by
\begin{align}
 M(c) = 1 \quad\mbox{and}\quad f(c) = \frac{1}{4}c^4 - \frac{1}{2}c^2 + \mu \zeta(\mathbf{x}) c,
\end{align}
respectively.
The local free energy $f(c)$ is modelled as a double well potential with two local minima corresponding to the two phases of the surfactant layer, similar to \eqref{gforLB}. The linear term introduces a tilt of the potential favouring the high concentration phase that is proportional to a spatially inhomogeneous term $\zeta(\mathbf{x})$, which parametrically mimics the shape of the static liquid meniscus occurring in the full model \cite{koepf2012substrate}. The resulting reduced model is considerably less complex, but still captures the essential physical aspects of the full model. This can easily be seen by direct numerical simulations of the model, see Fig.~\ref{pic-CH2D-homogeneous}. The resulting patterns are intriguingly similar to the ones obtained with the full model, which in turn clarifies the origin of the pattern formation process to be the phase separation dynamics of the surfactant monolayer, while the hydrodynamics of the underlying liquid layer is secondary.

In addition to direct numerical simulations, the simplicity of the reduced model allows for a more thorough analysis. A detailed bifurcation analysis of stationary states of the model was presented in \cite{koepf2012substrate}, also identifying the underlying local and global bifurcations that trigger the pattern formation process. The emergence of the entire bifurcation structure itself was furthermore investigated in \cite{KoTh2014n}.

\begin{figure}[htb]
\centerline{\includegraphics[width=0.8\textwidth]{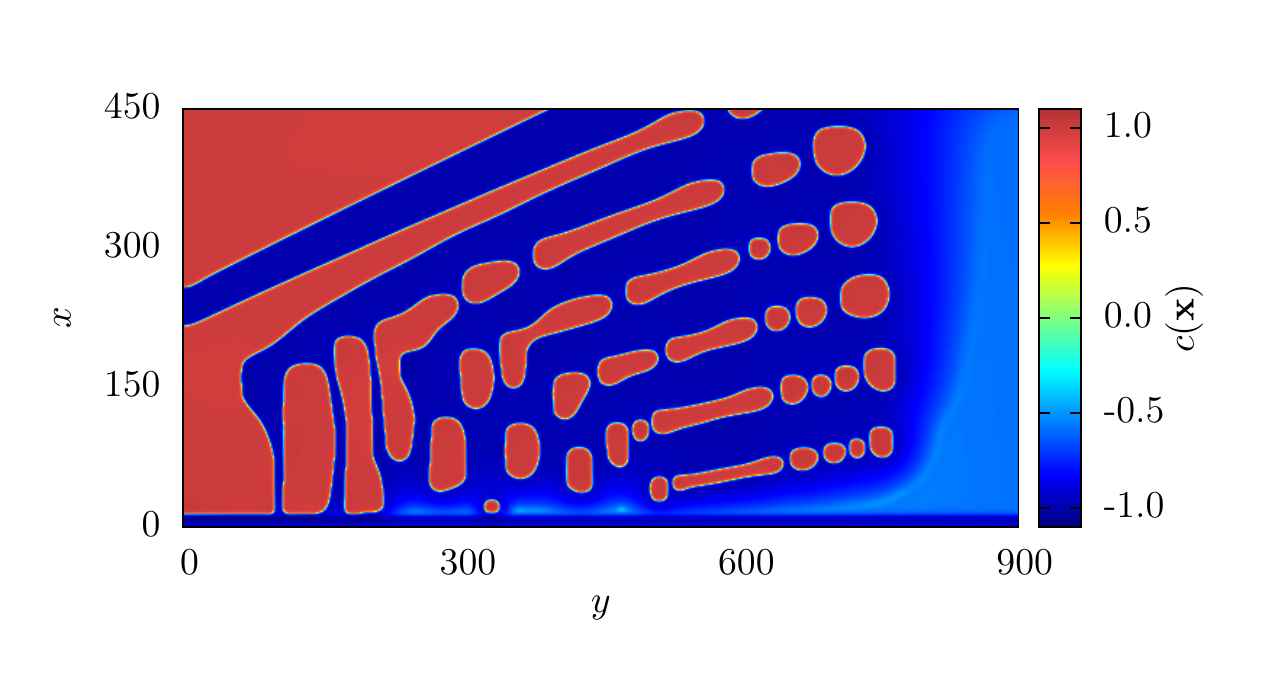}}
\caption{Snapshot of a direct numerical simulation of the reduced Cahn-Hilliard model for rotating Langmuir-Blodgett transfer \eqref{equ6} modelled by using an advection velocity field $\mathbf{v}(x,y)$ that corresponds to a rotation of the substrate around a centre at $(x_\mathrm{c},y_\mathrm{c}) = (0,-500)$ with a rotation frequency of $\omega = 4.6\cdot 10^{-5}$, leading to different effective transfer velocities across the substrate. As a result, the pattern type varies across the substrate from (i) a homogeneous high density area to (ii) stripes perpendicular to the meniscus to (iii) stripes parallel to the meniscus to (iv) a homogeneous low density area \cite{Wilczek2012msc}. Due to the strong velocity gradients, the stripe patterns tend to break up into droplets aligned on a line. For further simulation details and parameters see \cite{Wilczek2012msc}.}
 \label{pic-CH2D-rot}
\end{figure}

The reduced model lends itself to the introduction of additional effects, e.g., an interaction of the pattern deposition with a substrate prestructuring. As in the case of the two-component model for LB transfer \cite{koepf2011controlled}, such an interaction can lead to the production of more complex pattern types and to the occurrence of locking effects between the periodic prestructure and the resulting patterns \cite{Wilczek2014Locking}.

Besides modifications of the free energy, also changes of the transfer geometry and process can easily be incorporated, e.g., one may consider the so-called rotating transfer, where the substrate is not drawn out of the trough in a straight way, but instead with a rotational movement \cite{chen2007fabrication}. With this technique, different parts of the substrate are withdrawn at different transfer velocities, leading to a non-constant $\mathbf{v}(\mathbf{x})$. As the pattern formation process is strongly influenced by the transfer velocity, this yields complex patterns such as the one shown in Fig.~\ref{pic-CH2D-rot}, which can be understood as a sequence of the basic pattern types presented in Fig.~\ref{pic-CH2D-homogeneous} that all exist on a single substrate \cite{Wilczek2012msc}.

\section{Conclusion}
\label{sec:conc}
We have briefly reviewed three models that describe pattern formation phenomena in experiments, where complex fluids are transferred onto a substrate producing patterned ultrathin deposit layers with structure length from hundreds of nanometres to tens of micrometres. All models have been presented within the framework of a gradient dynamics formulation that clarifies how the various dynamics are governed by particular free energy functionals. These may be seen as extensions of the interface Hamiltonians for simple liquids. While the presented two-field models can be derived from basic hydrodynamic equations using a thin film approximation, we emphasise that the re-formulation as a gradient dynamics enables one on the one hand to easily compare and relate the different models to each other. On the other hand it allows for a direct and consistent way to extend the models by amending the individual contributions of the free energy and by including additional ones that account for further interactions and 
effects. It 
also offers a way to classify all such long-wave models derived from hydrodynamics based on whether they can be formulated as gradient dynamics or not.

Finally, we would like to highlight some open issues and relate the presented models to selected literature results in the wider field. Inspecting Eqs.~\eqref{equ5} with \eqref{equ5qq} and Eqs.~\eqref{VarForHPsi1} with \eqref{VarForHPsi1qq}, one might even see these formulations as 'trivial' because they have exactly the form that is expected for the evolution of two scalar conserved order parameter fields in the context of linear nonequilibrium thermodynamics \cite{Onsa1931pr,Onsa1931prb}. However, for two-field long-wave hydrodynamic equations this had to our knowledge  only been shown before for dewetting two-layer films \cite{PBMT04,PBMT05} but neither for surfactant-covered films nor for films of solutions or suspensions. Recently, the gradient dynamics form in the case of two-layer films has
been extended to include the case with slip and has been employed in deeper mathematical analyses of such models \cite{JHKP2013sjam,JPMW2014jem,JaKT2014cms}. We expect that the here reviewed gradient dynamics form will also facilitate similar analyses for films of suspensions, solutions and mixtures.

Note that the request that similar models for films of complex fluids in relaxational situations (without additional external driving forces and without inertia effects) should have a similar gradient dynamics form provides one with a good test for model consistence that a number of ad-hoc amendments introduced in literature models do not pass. It also provides one with a guide how to proceed in the further development of multi-field models, e.g., it is an interesting question how long-wave models for surface-active solutes and soluble surfactants \cite{warner2003surface,CrMa2009rmp} can be brought into gradient dynamics form even in the low concentration limit. The next step could then be to introduce into the relaxational gradient dynamics models additional nonequilibrium driving forces
that break the gradient dynamics form in a controlled way - an example are liquid films covered by a carpet of self-propelled surfactant particles \cite{AlMi2009pre,PoTS2014pre}.

Another interesting aspect is that the presented gradient dynamics form brings the long-wave hydrodynamic models into the context of models known as dynamical density functional theories (DDFT) for the time evolution of the density field(s) of (colloidal) fluids. They represent (normally diffusive) gradient dynamics on equilibrium free energy functionals derived with the methods of statistical physics \cite{MaTa1999jcp,ArEv2004jcp}, i.e., their main underlying assumption is that the two-point correlation function out of equilibrium can be represented by its equilibrium variant \cite{MaTa1999jcp}. This implies that the evolution equation for one and two fields is of the form of the present Eq.~(\ref{eq:CH}) and Eqs.~(\ref{equ3}), respectively.  However, due to the character of the energy functionals, the evolution equations are often integro-differential equations instead of the partial differential equations encountered here.  Note that in the two-field DDFT introduced in \cite{TVAR2009jpm,RoAT2011jpcm}, the 
diffusive mobility matrix is normally diagonal. How this is related to the choice of frame of reference for diffusion is discussed in the final part of \cite{XuTQ2015preprint}. It will be interesting to further pursue consequences of the formal similarity of DDFT and thin-film hydrodynamics, e.g., to develop models that combine the various pathways of transport.

\vspace*{0.5cm}

\section*{Acknowledgements}
This work was partly supported by the Deutsche Forschungsgemeinschaft in the framework of the Sino-German Collaborative Research Centre TRR~61. MHK was supported by LabEX ENS-ICFP: ANR-10-LABX-0010/ANR-10-IDEX-0001-02 PSL*.


\bibliographystyle{plain}
\bibliography{WTGK2015mmnp.bib}
\end{document}